\documentstyle[12pt]{article}

\def\'#1{{\accent19\ifx #1i \i\else #1\fi}}

\def\be{\begin{equation}}
\def\ee{\end{equation}}
\def\bea{\begin{eqnarray}}
\def\eea{\end{eqnarray}}

\newbox\Ancha
\catcode`@=11
\newdimen\ex@
\title{Standard-model coupling constants from compositeness}

\author{J. Besprosvany}

\date{Instituto de F\'{\i}sica, Universidad Nacional Aut\'onoma de M\'exico,
Apartado Postal 20-364, M\'exico 01000, D. F., M\'exico }

\begin{document}

\maketitle









\jot = 1.5ex
\def\baselinestretch{2.}
\parskip 5pt plus 1pt

\begin{abstract}

A coupling-constant definition is given  based on
 the compositeness property of some particle
states with respect to the   elementary states of other particles.
It is applied in the context of the vector-spin-1/2-particle
interaction vertices of a field theory, and the standard model.
The definition reproduces Weinberg's angle in a grand-unified
theory. One obtains coupling values close to the experimental ones
for appropriate configurations of the standard-model vector
particles,
 at the
unification scale within grand-unified models, and at the
electroweak breaking scale.

\end{abstract}


 \vskip 1cm

\baselineskip 22 pt\vfil\eject \noindent

The coupling constants are the dimensionless numbers that measure
the strength of nature's interactions.  Their values are fixed by
experiment in the standard model (SM) of elementary particles, and
depend on the energy scale. Clues to the origin of their values
are suggested from the relations among the quantum numbers of the
SM particles.

In general, the realization of unity among physical variables,
originally thought as disconnected, has led to a  new
understanding and connections among additional ones. For example,
by linking electric and magnetic phenomena, Maxwell's theory
showed that light is a  phenomenon of the kind, and predicted its
velocity in terms of likewise parameters. Indeed,
 recently proposed SM extensions including a unifying principle
are able to provide information on the coupling constant values.
Thus, grand unification$\cite{unification}$ assumes that the gauge
groups describing the interactions originate in
 a common group,
 and  it predicts a  single unified coupling, to  which distinct couplings indeed appear  to
 converge at high energy. It is also able to predict
the coupling-constant ratios. In addition, compactification
configurations of additional dimensions associated to
interactions\cite{Weinbergcoup},
 and  the dilaton-field  ground state
  in string theory\cite{Green}  predict their values, but, as yet, not
uniquely. Information on the coupling constants may be also
derived from extended-spin models\cite{Jaime}. Even if the
underlying dynamics is not obvious, these connections may become
manifest through symmetry arguments, which give additional
information.


Composite models are another class of unifying theories that
address the SM particle-multiplicity problem.   Utilizing the
connections among the quantum numbers of the 27 or so SM
particles, these particles are constructed in terms of fewer
elementary fields\cite{haplon}. The SM Poincar\'e symmetry and
gauge-invariant interactions provide the link.

In general, these symmetries dictate the few quantum numbers that
describe a particle state. These are the configuration or momentum
coordinates, the spin, the gauge-group representation, and the
flavor for quarks and leptons. Flavor characterizes only fermions.
In the SM, fermions belong in the spin-1/2 Lorentz representation,
and the gauge bosons are vectors. Similarly, fermions belong in
the fundamental representation of the gauge group, while the
vector bosons belong in the adjoint. This means  that the gauge
and spin quantum numbers of the latter can  be constructed in
terms of the former.

In the case of composite models, this facilitates their modelling
in terms of simpler fields. However, it is difficult then to
reproduce the SM dynamics without introducing additional fields
and interactions, which, in turn,  reduces the models'
predictability. Also, no additional substructure of the SM
particles has been found. Another appealing idea is to assume that
the vector bosons are composed of the SM fermions. A quantum
electrodynamics model was proposed in which the photon is
constructed from an electron and a positron\cite{Bjorken}. This
model requires an unobservable space asymmetry, and its
renormalizability rules are unclear.


In this paper, we use  the experimentally derived compositeness
property of
 the SM particles to  get
information on the SM coupling constants.   We focus on those
vector quantum numbers that  can be constructed in terms of those
of the  fermions. This is a remarkable SM property;  fermions
could otherwise belong to other   representations transforming
according to the Lorentz and gauge groups,
 without satisfying this property. As with grand unification, which
assumes a connection among the quantum numbers of the vector
bosons,
 this paper assumes a connection among
those of the spin-1/2 particles and vector bosons. The associated
symmetry provides the coupling information. In particular, the
application of quantum mechanical rules leads to normalization
constants, and Clebsch-Gordan coefficients that relate  both
representations, and
 ultimately relate to the coupling
constants. We will also find that the grand-unified coupling ratio
prescription is reproduced.

 In addition, we show that this assumption is
consistent with the SM. Indeed,  we apply an equivalent
field-theory formulation that makes this kind of compositeness
explicit, keeping the SM assumption that the fields are
fundamental, unlike the composite-model case; all the SM
predictions are therefore maintained. Thus, while composite models
require additional fields in terms of which SM or new particles
are constructed,  this assumption is model independent. Hence, the
putative problems associated with substructure compositeness are
not encountered.

We   first give a general coupling-constant definition  
based on  the normalization and the compositeness property of some
particle states with respect to other particle elementary states.
Using   the Wigner spinor classification of Lorentz
representations, one may express SM fields in terms of their
spinor components. It follows that the SM Lagrangian and its
fields can be rewritten and reinterpreted in this way. Finally, we
classify the configurations of the vector particles in relation to
their SM and grand-unified theory content, calculate corresponding
coupling values at the electroweak breaking  and unification
scales, and present  final comments.


Quantum numbers characterize particles, and the normalized state
$|w_i \rangle$ represents a particle with eigenvalue $w_i$ of the
appropriate  operator. The numbers $a_{ij}$ in the  composite
state
\begin{eqnarray} \label  {composite}
 | W \rangle =\frac{1}{\sqrt{N}}\sum_{i,j} a_{ij} |w_i \rangle|
w_j  \rangle  ,
\end{eqnarray}
 normalized with
\begin{eqnarray}
\label  {normalisation}
 N=\sum_{i,j} a_{ij}^* a_{ij},
\end{eqnarray}
fix  $\langle w_i w_j| W \rangle$. The same amplitude is
reproduced  by the corresponding operator
   $\hat  W=\frac{1}{\sqrt{N}}\sum_{i,j} a_{ij}
|w_i \rangle\langle w_j | $, satisfying $tr \hat W^\dagger \hat
W=1$, through $\langle w_i |\hat W |w_j\rangle$. Thus, both
structures keep the same information, and the same normalization
prescription may be applied.

$\hat W$ is also the most general operator acting on the $|w_i
\rangle$ states. Symmetry can determine the coefficients
$a_{ij}^\lambda$, up to a constant, where $\lambda$ labels the
 representation components of such symmetry. For example, the only (non-axial) vector
operator that can be constructed out of spin-1/2 particle states
is the Dirac matrix $\gamma_0\gamma^\mu$\cite{Dirac};
$\partial^\mu$ stems from  configuration space, and, when coupled
to a vector field, it is not relevant in the SM vector-spin-1/2
interaction Lagrangian because it is  neither renormalizable nor
gauge invariant. For each $\mu$ (no sum) $tr \gamma_0\gamma^\mu
\gamma_0\gamma^\mu=4$ normalizes covariantly the operator, and
fully determines it by providing the remaining constant; so is the
case for the  corresponding composite state $| W \rangle$. Hence,
the matrix element between the spin states $|i \rangle$ and  $|j
\rangle$
\begin{eqnarray}
\label{matrixelement}
 \langle i |\hat W^\mu|j \rangle
\end{eqnarray}
 is determined with $\hat
 W^\mu=\frac{1}{2}\gamma_0\gamma^\mu$. The four-entry $\hat W^\mu$
   acts on the space spanned essentially by the spin-1/2 particle,  its antiparticle, and
  their two spin polarizations.

This procedure can be generalized  to the case of  greater number
of degrees of freedom, using the rules for the direct product of
vector spaces and the generalized  operator  that acts on such a
space. The normalization   for $M$ such operators, $\hat W^T= \hat
W_1... \hat W_M$, is the product of  the traces of each operator
$\hat W_i$ in its space.

 The vertex interaction Lagrangian  $\int {\mathcal   L}_{f}$ with density $ {\mathcal   L}_{f}=-\frac{1}{2}gA^a_\mu{\Psi^\alpha}^\dagger\gamma_0
  \gamma^\mu G^a\Psi^\alpha$ is determined from  Poincar\'e and gauge invariance. In general, the latter determines the interactions of the
vector bosons with  the other particles, and among themselves, up
to the coupling constant $g$.    In particular, $ {\mathcal
L}_{f}$ is the only boson-spin-1/2 vertex. In the SM the fermions
belong in the fundamental representation. The vertex can be
consistently viewed as the expectation value of the tensor-product
operator
 $\hat W^{\mu a}=g \gamma_0\gamma^\mu G^a1_x 1_\alpha, $ with
vector components $A^a_\mu(x), $ acting upon the spin-1/2
particles $\Psi^\alpha(x)$;   $\mu$ is the spin-1 index, $G^a$ the
gauge-group representation matrix of the fermions,  $a$ the
group-representation index,  $x$ the spacetime coordinate  with
the diagonal\footnote{ $1_x$ only connects local fields, without
compositeness. Formally, $a^x_{x^\prime x^{\prime
\prime}}=\delta_{x^\prime x^{\prime\prime}}\delta_{x x^\prime}.$
$A^a_\mu(x)$ normalizes in $x$ space for $tr1_x=1$.}
$1_x=|x\rangle \langle x |$, and $1_\alpha$ the unit matrix over
the flavor $\alpha$. A composite state $A^a_\mu(x)|x\rangle |\mu
\rangle | a\rangle$,  with $|\mu \rangle$, $| a\rangle$  elements
as in Eq. \ref{composite}, underlies the operator association
leading to $\hat W^{\mu a}$:
 $|\mu \rangle\rightarrow (\gamma_0\gamma^\mu)_{\sigma\eta},$ $|a
\rangle\rightarrow G^a_{bc}$,  $|x \rangle\rightarrow|x\rangle
\langle x| $;   the   fermion state is $\Psi_{\eta c}^\alpha(x)$.
All are written explicitly in terms of  $\sigma$, $\eta$ spin-1/2
indices, $b,$ $c,$  gauge-group
 representation indices,  and the flavor. $A_\mu^a\hat W^{\mu a}$ is also the expression for the  vector field in
spin space, treated, e. g., in Ref.  \cite{Wald} (the same
generalization is applied to the gauge degrees of freedom). In
that reference, a spinor description of the Lorentz
representations is given. At each spacetime point, tensor
spinorial objects are defined. In particular,  a real basis of
(bi)spinorial objects is constructed that spans the Lorentz vector
representation. The component elements of such a basis are
essentially constructed out of the unit and the Pauli matrices. A
map is defined between these bispinor objects and vectors.
 Their  identification  follows from the  fact that
  they have the same transformation properties. In fact, Maxwell's equations can  be
equivalently formulated in terms of
 such objects, as two Dirac equations\cite{Bargmann}.
The other
 Lagrangian terms can also be reinterpreted and formulated in
 terms of
spin-projected fields.

Canonical quantization in quantum field theory  normalizes
$A^a_\mu$; the compositeness assumption further imposes such
condition on the  $\hat W_i$ operators, which fully normalizes
$A^a_\mu\hat W^{\mu a} $. In general,  $A^a_\mu $ can be
understood as an element in a polarization or group basis $A^a_\mu
=tr
 n_{\mu }^a  A^b_\nu  n^{\nu b}$, where  in our case $n^{\nu b }=\hat
 W^{\nu b}$, and it is assumed to be normalized. Indeed,  we recognize in
the vertex
\begin{eqnarray}
\label  {vertex}{\mathcal L}_{f}=-{\Psi_{\sigma
b}^\alpha(x)}^\dagger A^a_\mu(x)\Psi_{\eta c}^\alpha(x) \langle
\sigma |\gamma_0\gamma^\mu | \eta  \rangle          \frac{1}{2}g
\langle b | G^a | c   \rangle
\end{eqnarray}  the matrix elements in Eq. \ref{matrixelement}, and
the gauge-group ones.
 Within the compositeness assumption, we  equate each
matrix element in Eq. \ref{vertex} with that  of the composite
vector in Eq. \ref{matrixelement}, and similarly for the
group-representation matrices, all of which contain operators
acting upon the spin-1/2 particles, which  leads  to the
identification
 \begin{eqnarray}
\label  {identi}
  g\rightarrow 2\sqrt{ \frac{1}{ N}}.
\end{eqnarray}
 The    normalization  $N$ is
calculated as in Eq. \ref{normalisation},
 with the convention for the $\gamma$-matrices
\begin{eqnarray}
\label  {groupnorgam}
 tr  \gamma_\mu\gamma_\nu=4 g_{\mu\nu} ,
\end{eqnarray}
 and irreducible representations
\begin{eqnarray}
\label  {groupnor}
 trG_i G_j=2 \delta_{ij}.
 \end{eqnarray}
  Essentially, we are setting normalization constants for the
matrix elements in Eq. \ref{vertex}, which connect
representations, and can be viewed as Clebsch-Gordan coefficients.
${\mathcal L}_{f}$ contains sums over
 matrix elements for each $\mu$ and $a$, which  determine the coupling
constant; only two polarizations $\mu$ have a physical-state
interpretation, while gauge and Lorentz invariance demand a unique
value. Quantum field theory admits arbitrary coupling constants
for a vertex, which are obtained experimentally. The theoretical
assignment of $g$ complements this theory.

 In comparing the fermion states with
the vector ones, we find that the latter are
  composite only in the Lorentz and the gauge groups, whereas the configuration variable $x$ is elementary for both
  types
 of field.
  In general, an additional fermion index $\beta$ independent of $A^a_\mu$ corresponds to
  $\hat W_F=\sum |\beta \rangle\langle \beta| =1_F$, a unit
  operator present in the vertex,
 not contributing to the coupling constant. This is the  flavor's case.
 However, there are
 two consistent coupling definitions when such a kind of operator acts in a
 fermion subspace. Thus, e.g.,  $ SU(2)_L$  generators in a grand-unified theory such as
 $SU(5)$ are constructed
 with  their lepton  ($l$) and baryon ($b$) components as
$G_{SU(2)_Ll}+(G_{SU(2)_Lb}\times 1_{SU(3)})$, with $1_{SU(3)}$ a
projection operator in color space (leptons are color singlets);
$1_{SU(3)}$ does not commute with some $SU(5)$ generators,
 and the associated vector-field components interact
with the other unified-group ones. Physically, this  {\it full}
case corresponds to active
  degrees of freedom. In a lower energy regime, the symmetry is broken, and the  interactions are
truncated to the weak $SU(2)_L$   and the other  SM interactions,
while $1_{SU(3)}$ commutes with these generators. Then, in this
{\it reduced} case, $1_{SU(3)}$ drops out  of the calculation.


  Grand-unified theory predicts
coupling-constant
 ratios under the condition that the SM
 generators
 belong to the same unified-group representation, which determines Weinberg's angle at
 the unification energy scale\cite{unification}, and the running of
 the
 coupling of each interaction gives values at lower energies.

Similarly, the configuration  of  the fields'  group
representations $G_i$  gives a clue to the energy scale.
 To  obtain   unified and SM couplings
we specify the  normalized  vector-field polarizations and
gauge-group generators.  The couplings are calculated using the
fermion quantum numbers, which are the generators' eigenvalues,
and make the generators themselves (the Cartan subset).
 A generation of  SM left-handed   [quarks; leptons]  is classified
by $[Q,u^c,d^c;$  $L,e^c]$,  with  $L=(e,\nu)$, $Q=(u,d)$
$SU(2)_L$  doublets, and $u^c$, $d^c$, $e^c$,
 charge-conjugate singlets,
        according to their
 color-weak-hypercharge
 $SU(3)\times SU(2)_L\times U(1)_Y$ groups; the latter can be  viewed as subgroups of the $SU(5)$ grand-unified theory.
  The      multiplets are
 $ [( 3,2,1/3), (\bar 3,1,-4/3),(\bar 3,1,2/3);$ $(1,2,-1),
 (1,1,2)]$.
   The fermions fit neatly into the $\bf 5$ and $\bf 10$
representations of this group. The hypercharge $Y$ and the weak
interaction  have different $\sigma_{\mu\pm}=\frac{1}{2}(1\pm
\gamma_5)\gamma_0\gamma_\mu$ components,  with the pseudoscalar
$\gamma_5=-i\gamma_0\gamma_1\gamma_2\gamma_3$, which uses
${\mathcal L}_{f}$ with possibly different $\hat
W^a_{\mu\pm}=\sigma_{\mu\pm}G^a_\pm  $ components, in an obvious
notation.

 One gets for $Y$ in the {\it full} configuration,  with the above quantum numbers, the rules in
 Eqs.
 \ref{normalisation} and \ref{identi}, and   conventions in Eqs. \ref{groupnorgam}  and
\ref{groupnor},
 $g^\prime=2/[{2
(2 + 2^2 + 6(\frac{1}{3})^2+   3 (\frac{2}{3})^2 +3
(\frac{4}{3})^2)]^{1/2}}$ $=\frac{1}{2}\sqrt{\frac{3}{5}}$, where
the 2 in the denominator normalizes each chiral component
$\sigma_{\mu\pm}$,  to which corresponds one massless fermion
polarization.  The first two terms in the parenthesis
$2+2^2=1^2+1^2+2^2$ are the lepton hypercharges and, the  last
three are the quark hypercharges; their multiplicity is taken into
account.  $Y$ may be also viewed as a generator of the $SU(5)$
interaction.

Two  coupling definitions apply  for the
 weak $SU(2)_L$ interaction, one of whose generators has  diagonal components
  $I_{(l,b)} =(1,-1)$. For the {\it full}
  configuration, $g^{uni}=$ $2/{[2(1+3)(1^2+1^2)]^{1/2}}$
$ =\frac{1}{2}$, where the second factor in the denominator counts
the lepton and quark doublets, which in turn give the third
factor.
 Non-supersymmetric unified models\cite{DGLee}
 give  experimentally consistent unification
couplings of $g_{ex}^{uni}\sim.52 -.56$, at $10^{14}-10^{16}$ GeV.
From the SM\cite{Glashow}-\cite{Salam},
$tan(\theta_W)=g^\prime/g^{uni}$, and
 one reproduces the $SU(5)$ unification
result for Weinberg's angle$\cite{quinn}$
$sin^2(\theta_W^{uni})=3/8$. In general, the coupling definition
in Eq. \ref{identi} is consistent with the grand-unified
prescription for such a coupling ratio.

 The {\it reduced}  configuration of the normalized
weak vector   implies that  the color components drop from the
calculation.
 It gives the same
weight to quarks as to leptons, as is necessary if one omits
unification-group information. We should get information on the
electroweak-breaking scale to the extent that these weak and
hypercharge configurations describe on-shell Z and W vector
bosons. We find $g^{le}  =2 /[{2 (2)(1^2+1^2)
)]^{1/2}}=\frac{1}{\sqrt{2}}\approx
 .707$, while at the $M_Z$
 scale$\cite{tables}$, $g_{ex}=.649519(20)$,  where
one standard-deviation  uncertainty for  the last digits is given
in parenthesis.

 Each
isospin doublet component corresponds to a different hypercharge
isospin singlet;
this suggests, extending the rule to color components, that only
the {\it full}  configuration need be considered for $Y.$ Thus,
 $g'\approx .387$ is between $g_{ex}^\prime(M_Z)=.35603(6)$
 and the unified hypercharge values $\sqrt{\frac{3}{5}}g_{ex}^{uni}\sim .40-.43$;
the relatively narrow range provides a
 test of the prediction.  From $tan(\theta_W)=g^\prime/g^{le}$,
 we find $sin^2(\theta_W)=3/13\approx .23078,$
 while at $M_Z$ $sin^2(\theta_{Wex})= .23113(15).$


    One may also interpret  the {\it reduced} weak configuration
    within the minimal supersymmetric
 model, with a unified\cite{Amaldi} $g^{Suni}_{ex}=.69(4)$ at $10^{15.8\pm  .4}$ GeV. $g^{le}$ reproduces
a value also within a narrow low and high energy range.
 For the gluons' coupling in the $1_{SU(2)_L}$-{\it reduced} case we use the $\lambda_3$ Pauli-matrix  fundamental
component of the $SU(3)$    (any other generator would also  do)
with the convention of Eq. \ref{groupnor} $g_s =2 /[{2
(2)(1^2+1^2) )]^{1/2}}=1/\sqrt{2}\approx
 .707,$ or $\alpha_s=\frac{g_s^2}{4
\pi}\approx .040$,
 while $\alpha_{s(ex)}(M_Z)=.1172(20)$.
 Then $g_s$ provides a  lower limit around the unification scale.

While only  the fermion-vector vertex has been examined, the
results are valid for a more general Lagrangian.  The coupling
constants in the other Lagrangian terms get a unique value,
because gauge invariance demands it for each gauge group.
 All along, flavor is
assumed to belong to the {\it reduced}  configuration, for it does
not influence interactions.

In a grand-unified theory and  in the SM, the electroweak-field
components at the  unification  scale,
 and  at the symmetry-breaking scale,
are determined, respectively, through the ratios of the
electroweak couplings, namely, Weinberg's angle. The     SM
fermions and bosons, and their simple interactions conform to  a
compositeness assumption. Under this assumption, the allowed
vertices and the fields' normalized polarization generate the
coupling constants. Specifically, these are obtained by
associating composite-field configurations both to  the
unification scale and
 the  W and Z particle regime. Already at tree level,
Weinberg's angle is reproduced for the $SU(5)$ unified theory, and
a value close to the experimental one is obtained at the $M_Z$
electroweak-breaking scale,
 which
validates
 the ascribed configuration in each regime.
This  set of two coupling constant or Weinberg angle values
provides a connection between the two energy scales through, e.g.,
the renormalization group equations, which have to be supplemented
with boundary conditions.
Although the low-energy  ratio $tan (\theta_{W}) $ does not
contain couplings at  precisely the same energy, it contains
information on the group-generator structure, stemming from the
compositeness assumption; this is not in contradiction with the
coupling running that should be applied, and whose corrections
cancel among the two couplings. The couplings are also interpreted
consistently  within the minimal supersymmetric model. The
calculation of $\theta_{W}$ can be viewed   as   complement to, or
as alternative to, that of $\theta_{W}^{uni}$. In the first
approach, the coupling constants relate energies in the unified
and symmetry-breaking scales.
 In the second approach, one obtains information on the $M_Z$
scale, understood as fundamental\cite{Dim}.

The  paper's approach may also be applied in other extensions,
which require only the consideration of reducible representations.
The compositeness hypothesis is supported  with coupling constants
obtained  among a limited number
 of allowed configurations, and that  reproduce experimental
 values,
 which are within a narrow range at different energy scales.








\setcounter{equation}{0}


{\bf Acknowledgement.} The author acknowledges support from
DGAPA-UNAM,   project IN120602, and CONACYT, project 42026-F.









\end{document}